# A New Case For an Eternally Old Infinite Universe

© Abhas Mitra

Theoretical Astrophysics Section, BARC, Mumbai-400085, India
Email: amitra@barc.gov.in

**Abstract:** We start with a new version of Newtonian cosmology by incorporating the special relativistic fact that the galaxies are losing mass due to emission of radiation. This simple fact yields accelerated recession for the galaxies in accordance with observations. We point out that in the presence of accelerated expansion, the universe can be infinitely old and suggest that the observable universe is only a speckle of the true universe. From various considerations, we argue that the mean density of the universe must be zero and the cosmic fluid comprises infinite number of such speckles separated by infinite distances. The Microwave Background radiation is shown to be just the sum of redshifted thermal radiation of Eternally Collapsing Objects (ECO), the so-called Black Hole Candidates. The hot photosphere of the ECOs cooks light elements the same way they are supposed to be produced in hot early universe. We indicate how White Dwarfs, Brown Dwarfs, Stars, Neutron Stars, ECOs all accrete and wither away by gravitational collapse to merge into the interstellar medium only to be reborn to keep cosmic machine working even in an infinitely old universe. This universe with nested infinities has zero baryon number because it contains equal number of matter and antimatter ``atoms''. The predicted microwave luminosity of the galactic centre ECO (Sgr A$^*$) nicely matches with the corresponding WMAP estimate.

## 1. Introduction: challenges for standard cosmology

Although the Standard Big Bang Cosmology (SBBC) is definitely the most popular, successful and viable cosmology, recent WMAP results have thrown in several unexpected challenges for it apart from the preexisting fundamental problems of ``Fine Tuning'' and ``Particle Horizon''. The SBBC apparently overcame the latter two problems by invoking the rather unnatural hypothesis of ``Inflation''. And inflation, by ensuring homogeneity and isotropy, predicts that fluctuations of the so-called Cosmic Microwave Background Radiation (CMBR) must be Gaussian and entirely random. However, more and more WMAP observations are revealing that there are lot of patterns in the small fluctuations of supposed CMBR[1,2,3]. Moreover, at large angular scale in the sky, while there are regions where CMBR is smoother, there are regions where it is lumpier. One of the dramatic findings is the existence of a huge ``cold spot'' in the CMBR sky which has been attributed to a void of dimension 280 Mpc[4]! We may also note that CMBR sky hardly displays ``the shadow of Big Bang'', i.e., intervening old dense clouds of plasma hardly scatter the CMBR coming from behind (Sunyaev –Zeldovich effect)[5,6]. The detection of a WMAP ``haze'', i.e., significant excess of CMBR from an extended region around the galactic centre is also alarming even though attempts have been made to explain away this ``haze'' as a synchrotron emission of $e^-/e^+$ pairs resulting from decay of ``cold dark matter''[7].

SBBC predicts that the nuclide D is of complete primordial origin. If so, relative D abundance must be same everywhere. But in 1996, there was a report that the ratio of D/H is 3 times higher than the primordial value in some galaxy. If true, this report would imply local production of D in that particular example[8]. In contrast, it is known that He-4 is generated in stars in addition to likely primordial origin. If so, the the abundance of He-4 in stars can only be higher than the primordial value. But Luca Casagrande et al.[9] have shown that the old main sequence stars have much less He-4 than predicted by Big Bang nucleosynthesis! They found that the stars that have a metal < 1.3%, the average helium abundance is 18±2% well below the Big Bang value. This finding appears to be in contradiction with SBBC.

Then there is the old problem of ``Born Before Big Bang'' (BBBB). However, after the discovery of a supposed accelerated Hubble expansion, the present version of Cold Dark Matter- Λ SBBC apparently removes the puzzle of existence of stars in the globular clusters older than the age of the universe by a choice of [10] $\Omega_\Lambda$=0.7, $\Omega_{matter}$=0.3, H =60 Km/s/Mpc, where $\Omega= \rho/\rho_c$, ρ is the density of universe, $\rho_c$ is the critical density for closure, Λ indicates cosmological constant or Dark Energy parameter and H is the Hubble's constant. Although the problem of BBBB might be resolved in this context, the problem of galaxy and structure formation in earlier young universe may persist. For instance the universe at a redshift of Z=6.5 is already full of galaxies and rich structures even though it may be only 550 million years old[10]. Romani et al. detected the far away blazer Q0906+6930 at Z=5.74 which contains a BHC of mass M>$10^{10}$ M$_*$, where M$_*$ is solar mass. The existence of such a massive BHC at a time ``The universe awfully young'' is an unsolved mystery[11].

In general, there is growing evidence that the observed universe is far from smooth and homogeneous even though it might be crudely isotropic[10]. For the last 30 years, deep surveys have shown that galaxies and clusters are preferentially located within the walls of spongy or honeycomb like structures surrounding huge voids which is a far cry from the text book picture of ``smooth homogeneous'' distribution of ``cosmic fluid'' assumed by not only SBBC but by most of the other cosmologies too.

There is more or less an unanimity that the universe, as observed from Earth, the matter distribution has a fractal structure with index $D$=1.2 atleast upto $R_{max}$ =10 Mpc[10]. And many cosmologists think that fractal structure of matter distribution has been found upto $R_{max}$=100 Mpc ($D$=2.0)[10]. It may be reminded that even if matter is distributed uniformly filled regions of fractals, the net matter content would increase as $R^D$ whereas the same would increase as $R^3$ in a truly uniform distribution. We cannot rule out the possibility that future surveys would not show the observable universe to be of fractal nature upto largest *observable* scales with $D \leq 3$. How to address such severe problems confronting SBBC ? And how to address the general conflict between the observed fractal structures of the matter distribution with the idea a ``smooth and uniform'' ``perfect fluid'' description of cosmic matter inherent in most of the cosmologies?

To answer such questions, we develop below a new cosmological paradigm of an ``Eternally Evolving Infinite Universe'' (EEIU). Let us recall that there are elements of assumption, speculation and extrapolation in many scientific theories and much more so in cosmology because of its very nature. Many physical theories have been developed also from a regard for beauty, simplicity and symmetry. In particular, philosophical considerations have also guided framing of cosmology as a science; the belief of inevitability of ``isotropy'' and ``homogeneity'' has more to do with good philosophy rather than pure science per se. Such considerations too have played their due role in the formulation of EEIU.

## 2. Explaining Accelerated Expansion without Dark Energy

In Friedmann Robertson Walker (FRW) model of standard cosmology, the dynamic equation involving acceleration/deceleration of the observed universe is

$$\frac{\ddot{S}}{S} = -\frac{4\pi G}{3}(\rho + 3p/c^2) \tag{1}$$

where G is the Newtonian constant, c is the speed of light and S =S(t) is the cosmic scale factor, t is the cosmic time and an overdot means differentiation w.r.t. t. If the fixed comoving coordinate is r, the areal coordinate R(r,t) = r S(t), and one may rewrite Eq.(1) as

$$\frac{\ddot{R}}{R} = -\frac{4\pi G}{3}(\rho + 3p/c^2) \tag{2}$$

In order that $\ddot{S}$ >0, one must *modify the FRW model in a rather adhoc manner by* introducing a sufficiently large positive term on the RHS of Eq.(1) or (2). And this the reason for invoking ``Cosmological Constant'' or suitable ``Dark Energy'' in the FRW model. Recall that in all standard cosmologies the cosmic fluid is assumed to be not only smooth and uniform on all scales but it must be dissipationless and non-radiative too. In particular Weyl's postulate assumes the ``atom'' of the cosmic fluid to be a ``galaxy'' in free fall and which is not undergoing any kind of interaction with its surrounding other than through gravitation. But in reality, *all galaxies are radiating* and that is the reason we can see them. Note that this aspect is not all incorporated either in FRW or any other model. So, can we modify the FRW model to incorporate the basic property of galaxies?

Suppose we are considering the so-called ``flat model'' for which proper volume is M(R)=4πR³/3 and the enclosed mass is M(R)=4πρR³/3. In such a case, it is well known that, if we forget for a moment the tiny p/c² term in Eq. (2), it is nothing but the good old Newtonian Equation of motion of a test galaxy of mass m in the gravitational field of the enclosed mass:

$$F = -\frac{GMm}{R^2} \tag{3}$$

For ``closed'' or ``open'' models, Eq.(3) is valid only for sufficiently small ``r'' because the expression for proper volume would deviate from the ``flat'' Eucledian one. It was noted long ago by McCrea & Milne[12] that most of the basic equations of cosmology starting from Hubble's law can be obtained by purely Newtonian laws (except for p/$c^2$ like terms)! And if one would only incorporate the special relativistic concept of equivalence of mass energy associated with pressure (which is an energy density), simple Newtonian considerations would yield Eq.(2) which normally is obtained in ``Relativistic Cosmology'' after painstaking elaborate exercises in Tensor analysis and General Relativity [13]! Even now, there is no answer to this deep mystery and in the latter part of this paper we would address this issue. So, we fall back to Newtonian cosmology to seek an appropriate modification of Eq.(2) which would incorporate the fundamental physical fact that galaxies are radiating, say with a luminosity of

$$L = -\frac{d(mc^2)}{dt} = ml \quad (4)$$

where l is the luminosity per unit mass. The Newtonian cum special relativistic equation for energy conservation of the test galaxy is

$$mc^2 + \frac{1}{2}mv^2 - \frac{GMm}{R} = cons \quad (5)$$

Here $v = \dot{R} = H(t)R$ is the Hubble flow speed of the test galaxy. Since there is no *net energy flow* in a supposed homogeneous universe mass energy of any section M(R,t) remains fixed. Now if we differentiate the foregoing equation w.r.t. cosmic time t and drop terms of the order of $v^2/c^2$, we would obtain:

$$\ddot{R} = \frac{l}{v} - \frac{GM}{R^2} = \frac{l}{H(t)rS(t)} - (4/3)\pi G\rho(t)rS(t) \quad (6)$$

instead of Eq.(2). At a given t, the above equation shows that, small r regions would be dominated by the 1st term on the RHS. This means that, there should be an acceleration zone of radius R=$R_*$ given by

$$R_* = \left(\frac{3l}{4\pi GH\rho}\right)^{1/2} \quad (7)$$

and beyond which there should be deceleration due to the 2nd term on the RHS of Eq.(6). At the present epoch, Sun has l =$l_*$ = 2 erg/s/g, and if we take present value of H =$H_0$= 60 km/s/Mpc, and ρ=$\rho_c$, we find $R_*\approx$ 7x $10^{26}$ cm ≈ 240 Mpc and the corresponding $Z_0\approx$0.05. Present observations *explained in the framework of homogeneous FRW model,* however suggest that galaxies as far away as Z ~1 could be in a state of accelerated expansion. And if we depart from this model, and allow for likely fractal structure upto Z≈1, it might be possible that true Z of accelerating galaxies are smaller. In any case, this toy model generates acceleration in terms of transparent well known physical facts rather than by ad hoc mysterious non-verified ideas of ``Dark Energy'', ``Quintessence'' or ``Phantom Field'' etc.. Once such requirement of ``Dark Energy'' gets eliminated, one may think of a mean ρ<< $\rho_c$ and then Eq.(7) may yield an acceleration zone with $Z_0 \sim 1$. For a fractal universe, such a low value of ρ cannot be ruled out.

In the light of this exercise, the standard FRW Eq. (2) should be modified into:

$$\frac{\ddot{R}}{R} = \frac{l}{HR^2} - \frac{4\pi G}{3}(\rho + 3p/c^2) \quad (8)$$

**3. Big Bang or Big Whimper?**

The conventional claim about inevitability of Big-Bang singularity is made by analyzing Raychoudhuri equation[14]. Naturally, this analysis considers the cosmic fluid as perfect and non-dissipative. *Further it is assumed that there is no acceleration,* $\ddot{S}(0) \leq 0$. It is then found that volume expansion factor $\Theta = \infty$ at a finite proper time in the past. The age of the universe

$$t = \int_0^{S(t)} \frac{ds}{\dot{S}} \quad (9)$$

is bound to be finite for a finite $\dot{S}$ (t) because for Big Bang, $\dot{S}(0) > \dot{S}(t)$. Blowing up of expansion factor may also imply $H = \infty, \ddot{S} = \infty$ at t=0. One might see that such a singularity is inevitable because Big Bang assumes the *instantaneous* appearance of a finite size universe out of nowhere i.e., from a mathematical point of zero extent.

Such an initial condition is unphysical not only from the viewpoint of violation of all conservation laws but also from much simpler physical reasons. For a moment, let us consider the mundane case of a laboratory explosion of a ball lying at rest at t=0. In order for it to ``explode'' some external agent (heat, electricity, pressure) must release energy in a very short span δt, and the ``bang'' must necessarily begin with an acceleration rather than a deceleration. Only after this brief initial phase of acceleration, the explosion fragments may undergo deceleration either due to self-gravity or some dissipative process. Thus it is completely unphysical to conceive of a ``bang'' which starts with a deceleration. If it has to only decelerate ever since the very beginning (while v=0), why would it explode at all and not implode?

One might also see some basic inconsistency for such initial conditions: If one would indeed have $\dot{S}(0) = finite$, one would also have $\ddot{S}(0) = +\infty$ because after all everything was at rest before the birth. Then *how can there be a deceleration at the same moment, $\ddot{S}(0) \leq 0$ ?* And even if we accept that $\ddot{S}(0) \leq 0$ by ignoring the inherent contradiction, further inconsistencies in Big Bang initial conditions may be found:

At t=0, the scale factor S(0)=0. *Then how can S(t) acquire a finite value without $\ddot{S}$ being positive ever?* Instead, we feel that one must have initial $\dot{S} = 0$ both for mathematical continuity and physical consistency particularly in a case where no external agency is involved to trigger the expansion. To appreciate our contention, consider the fact that the universe might have been born out of some quantum fluctuation. Even in such a case, there must atleast be some finite duration δt after which a finite universe can appear from nowhere. And during this interval, there must be an acceleration from a state of rest, i.e., $\dot{S}(0) = 0$. Thus even if would accept that the Universe appeared from nowhere, Big Bang should really be a Big Whimper from simple physical considerations:
$$S(0) = 0; \dot{S}(0) = 0, \ddot{S}(0) \geq 0. \qquad (10)$$
In the presence of acceleration or rather in the absence of monotonous deceleration, Raychouhuri analysis gets invalidated, and Eq.(9) allows for solutions with $S(t) = \infty; ..t = \infty$ (in the present epoch) *even when* $\dot{S}(t)$ positive definite: which essentially requires infinitesimal positive acceleration. <u>The observations of an accelerating universe may be finally signaling that Big Bang was actually a Big Whimper in infinite past</u>.

In the context of Big-Bang, if a test particle would indeed be *always* governed by Eq.(3) and there is no external agency involved, and if neither is there any dissipation of any sort, it can never expand away unless somehow M decreases. On the other hand, in Eq.(8), the accelerating term can dominate over the decelerating term even if $\rho \to \infty, ..p \to \infty$ in the limit $t \to 0, H \to 0, R \to 0$. Hence Eq. (8) can indeed be consistent with the physical initial conditions of a Big Whimper.

## 4. Eternally Evolving Infinite Universe

Despite the inherent assumption of uniformity and isotropy on all measurable scales, the nearby universe comprising stars, planets, meteors, interstellar clouds, bubbles is grossly non-uniform and this is known from ancient times. Thus during the formulation of modern cosmology, it is the galaxies which were considered as ``atoms'' of the *non-dissipative, smooth and uniform cosmic fluid* of FRW model. With the improvement of observational prowess, subsequently, galaxy clusters or super clusters usurped this title. However, there was unexpected surprise when all luminous matter of the universe was found to be preferentially located in sheets and walls around huge voids. Then what is really the structure which might be taken as ``atom'' of the cosmic fluid? It has often been suggested that universe is nevertheless a ``smooth fluid'' over a scale of ~100 Mpc. But this problem gets unpleasantly complicated in view of the fractal like distribution of luminous matter on various scales with decreasing mean density. Let us again recall that the size of the largest void is 280 Mpc and it would be naïve to assume that matter distribution then could be smooth and uniform over a distance of say 2800 Mpc. And even if we accept such a scale of uniformity, it would be close to the Hubble Length c/H and the idea of a uniform matter distribution would be severely dented.

In addition, such inhomogeneities seen from Earth would challenge the Copernican principle because it might appear that Earth, the habitat of life, may indeed be located in a preferred position in the cosmos! To blunt such possibilities, we propose that the ``atom'' of the cosmic fluid is the entire observable (patch) universe. And this notion, by definition, is independent of the state of observational capabilities at any epoch. This model further requires that this radiating galaxies/stars are submerged in a non-luminous matter of infinite extent so that the net luminosity per unit mass of this dust $l = 0$. And this non-luminous baryonic matter could be in the form cold neutral hydrogen clouds of primordial origin and which replenishes the loss of ``mass'' into energy from the luminous matter.

If so, then what is really the ``whole universe''? Well, the universe comprises infinite such ``atoms'' of infinite extent, each separated from the neighbours by bigger infinite distances! So, this universe is an infinity of infinities! And even if the mass of a given dust atom would be infinite, the density of this infinitely diluted universe could be zero because it is possible to have $\infty/\infty = 0$! Having a mean density of zero and infinite size, such an universe is always homogeneous and isotropic and probably only such a model can satisfy the mathematical requirement of a perfectly isotropic and homogeneous perfect fluid required by various cosmologies. Accordingly there is no interaction between the particles of this infinitely diluted dust. Gravitational energy is zero everywhere except within the inner regions of the ``dust'' particles.
How did such an universe become ``isotropic'' and ``homogeneous'' where ``atoms'' are infinitely separated. The simple answer is that, the age of this universe is after all infinite.

Since $l = 0$, for these dust grains, and gravitational field is also zero, the ultimate universe is coasting freely with a deceleration parameter $q = 0$. This is possible if the infinitesimal acceleration eventually died down in infinite past. Some inner part of the patch consisting of luminous galaxies, however, is accelerating because of local dynamical effects ($l > 0$) in accordance with Eq.(8). As mentioned earlier, in the absence of any deceleration, Eq.(9) admits an infinite age. The scale size of the universe at two intervals $t_1=\infty$ and $t_2=\infty$ are $S_1=\infty$ and $S_2=\infty$ respectively even though $(t_2 - t_1)$ =finite and $(S_2 - S_1)$ =finite. Falling back to the popular picture of mutually receding dots on an expanding balloon, the *finite size dots too expand away in response to global expansion.*

*This model may explain WHY HUBBLE FLOW IS SO SMOOTH despite lots of inhomogeneities and peculiar motions in the galactic neighbourhood.* The regularity of the Hubble flow results not from uniformity or the lack of it of objects lying within the ``patch'' in which we live. On the other hand, it is due to uniformity of the distribution of the ``patches'' or ``cosmic atoms''. The Hubble flow within the patch is a result of the space expansion of the overall universe and the acceleration zone may extend to $Z_0\sim1$ because of sufficiently low value of $\rho_{patch}$. A question has been raised that creation of space by Hubble expansion entails violation of energy conservation because ``vacuum'' is endowed with ``mass-energy''[10]. On the other hand, it is most likely that quantum vacuum fluctuations exactly cancel each other to generate a zero vacuum energy density and a zero cosmological constant. If vacuum would indeed have a huge energy density, simple laboratory experiments should have detected it long back. We may note two variations of this model

(a) Infinite Static Model:

Here the cosmic fluid is not expanding at all and the universe is the original static Einstein universe with ρ=0 and Λ=0. There is no problem of its stability because both density and gravitation force are nil everywhere except within the ``atoms'' or observable ``patches''. The ``patches'' themselves must have sufficient low density and a fractal structure to avoid any stability problem. The accelerated expansion of a ``patch'' in this case will be due to dynamical effect of radiation loss of the galaxies and not necessarily due to any ``space expansion''. However, the problem of smoothness of the Hubble flow on Mpc or even larger scales would continue to remain a puzzle in this case.

(b) Truly Static Model:

Above models are built to accommodate the fact that as of now, the best explanation for observed Red Shift of distant galaxies appears to be of dynamical origin. However there are contending proposals like ``Tired Light Hypothesis''. More recently, there is a proposal that the plasma environment around quasars may result in intrinsic redshifts [15]. There is also an anomaly -- why the surface brightness of highly distant galaxies are more or less same as that of nearby galaxies. Although, it is quite improbable, imagine that in future, it would learnt that the observed redshifts are not of dynamical origin. Then, we must have truly static version of EEIU where the ``patches'' too are static like the background. The value of Hubble constant in this case would be zero on all scales. This would be a revision of Einstein's ``Static Universe'' which is of zero density inspite of possessing an infinite amount of matter. Philosophically, this would be the ideal universe.

**5. Why Zero Density?**

Let us ponder why the density of a *truly isotropic and homogeneous universe* should be zero. In General Relativity, in the presence of gravitation, usually two clocks at rest with respect to each other cannot be synchronized unless they are spatially coincident. And thus universal synchronization of clocks in the presence of gravity is against the spirit of GR. For isolated bodies, this however can be achieved if one would imagine a very unphysical fluid which has no pressure, p=0, despite having finite density, ρ>0. Such a fluid is termed as ``dust'' and is used only for the desperate need to obtain analytical solution of problems which are otherwise completely intractable. The proper time of collapse of a uniform dust ball to a singularity is:

$$\tau = \frac{\pi}{2}(3/8\pi G \rho_0)^{1/2} \qquad (11)$$

where $\rho_0$ is the density of the dust when it is at rest. But *a dust can never be at rest because of absence of pressure gradient force. However mathematically one may assume that $\rho_0 = 0$ and hence it can remain in hydrostatic equilibrium even in the absence of any pressure or its gradient.* Thus for physical consistency, one must have ρ=0 for a ``dust''. This is also expected from thermodynamics to ensure that sound speed with the fluid is finite. Accordingly one should have τ=∞ for dust collapse. Incidentally, the Newtonian formula for dust collapse <u>exactly coincides with the corresponding GR formula (11)</u>! Is it a mere coincidence? No. It is so because *only in the limit of $\rho \to 0$, GR exactly coincides with Newtonian gravitation*! We know that, in Newtonian physics, all clocks can be universally synchronized to yield an universal time. One does so for a `dust'' in GR too because inherently a ``dust'' like condition can be fulfilled only for ρ=0 [16,17,18]!

In contrast, the cosmic fluid in the FRW model *is supposed to possess* arbitrary high pressure apart from arbitrary high density. But there is no pressure gradient to support the fluid against its self-gravity! And in principle, there can be static solutions too even in the absence of any counterbalancing force on the fluid!! This has always been puzzling and has recently led to suggestion to modify GR itself 19]. And although there should be a strong grip of gravity for the fluid in view of likely arbitrary high density, far away clocks can be synchronized universally! Finally, as noted earlier, despite having arbitrary high pressure and density, the FRW evolution equations are exactly similar to their Newtonian counterparts except for additional terms like p/c². Note unlike the dust collapse case, FRW solutions are not supposed to be any approximations for physical realities, on the other hand, they are supposed to be fundamental exact solutions. The only answer to series of such conundrums is that the FRW solution is physically valid only for strict ρ=0 and p=0 case just like the dust solutions. There can be another less fundamental reason for a universe with ρ=0:

We know that the external Schwarzschild solution is valid even in a non-static case as long as there is no emission of radiation due to mass motion. And in order that peculiar speeds never attain the speed of light, there must not be any Black Hole like condition of $R = R_g = 2GM/c^2$.

If we momentarily assume an uniform density $\rho = \rho_c = 3H^2/8\pi G$, we would find that, we would indeed have this unwanted Black Hole like property at R=$R_g$= c/H.=Hubble Radius! For a smaller density, we would have $R_g$> c/H, and for a larger density, instead, we would have $R_g$ < c/H. Random motions and heat flow can mitigate such a catastrophe for a physical fluid. But for FRW fluid, there is no heat flow and random motions do not help either because there is no pressure gradient. To be more specific, it is known that, the Active Gravitational Mass Density (AGMD) of a spherically symmetric fluid is

$$\rho_{AGMD} = \sqrt{g_{00}}(\rho + 3p/c^2) \qquad (12)$$

where $g_{00}$ is the parameter responsible for synchronization of clocks. Since, in general, in the presence of gravitation, $g_{00} < 1$, the $\rho_{AGMD} < (\rho + 3p/c^2)$, contrary to what is often mentioned by many authors. Fur-

ther, Mitra [20,21] has shown that, actually

$$\rho_{AGMD} = (\rho - 3p/c^2) \tag{13}$$

But both for dust and FRW fluid, by fiat, $g_{00} \equiv 1$, and one artificially obtains the wrong notion that as if

$$\rho_{AGMD} = (\rho + 3p/c^2) \tag{14}$$

and pressure cannot help in reducing the effect of gravity. Thus for a truly FRW universe with uniform finite density, somewhere or other, the peculiar inward speed of the galaxies might become the speed of light! And by the principle of isotropy and homogeneity, then, the peculiar inward speed of galaxies everywhere could be equal to speed of light !! And a resolution to the apparent contradiction between Eqs.(12), (13) & (14) lies in the realization that mean value of ρ=p=0 for the cosmic fluid. Can a fractal model of the universe avoid this catastrophe? Yes, only if the fractal dimension $D$=1.0 on sufficiently large scale because M(R)/R would be constant in such a case. But even if one would accept that fractal model spans the entire observable luminous universe, it is seen that $D$ increases outward and probably approaches 3. Thus from this consideration too, the ultimate universe must have a mean ρ=0.

Finally recall that many authors have worked out the mass-energy of FRW universe by using various forms of gravitational stress-energy tensor by using not only GR but also tele-parallel gravity [22]. And almost in all the cases, one finds M=0. For a non-singular extended body, this is possible only if mean ρ=0. For a singular case, however, it is possible to have M=0 because $\rho_{AGMD} \to 0$ as $\rho \to 3p$ even if $\rho \to \infty$.

### 6. Evolution: Young Stars in an Old Universe?

A natural question for EEIU would be how could there still be shining stars in an infinitely old universe when stars have finite lifetime. We already know that stars not only die but many of them are still being hatched in interstellar clouds. Thus the above question would degenerate into the question of the existence of clouds in an infinitely old universe. The key to this question lies in the physical properties of Eternally Collapsing Objects (ECOs) which are always churning out not only fresh hydrogen but most of the lighter nuclides too. How do they do so?

An ECO of mass M has a mean local temperature of $T \approx 600 MeV (M/M_*)^{-1/2}$ where $M_*$ is the solar mass [23]. Thus while stellar mass ECOs are mostly pure energy and $e^{+/-}$ pairs, whatever baryons are there, they are in the form a quark gluon plasma (QGP). Above the core of an ECO, there is a photosphere from which ECO radiation leaks out; and above it there is a much cooler atmosphere. A stellar mass ECO is the realization of the mythical ``hot big bang'' like environment. Some region of the ECO photosphere is cool enough (~1 MeV) to cook light isotopes such as D, He-3, He-4, Li-7 etc. For a supermassive ECO with $M \sim 10^6 M_*$ (Sgr A$^*$), the mean T= 600 KeV and the entire ECO is ideal for light isotope production. More massive ECOs too can synthesize light elements at their hot inner cores. An ECO being supported almost completely by radiation pressure, rather than by gas pressure, is extremely vulnerable to radiative eruptions during which ECOs eject QGP and synthesized light nuclides into the ISM. Major ejection of QGP and overlying stellar material however occurs during the birth of ECOs in the form of Gamma Ray Bursts (GRB). In fact, during its formation, an ECO can undergo several massive outbursts (GRBs) in a matter of few days. Ideally, the life of an ECO is infinite if we ignore any loss of baryons from it. But in reality, ECOs keep on erupting in unpredictable manner because of radiation pressure and this may be reason that the so-called BHCs in X-ray binaries show much richer temporal variability than their Neutron Star (NS) counterparts. Even though ECOs keep on accreting from ISM, presumably, they eventually evaporate completely by returning all their QGP in the core and cooked H, D, He-3, He-4 in the envelope. Since the ECO is a practical realization of the mythical ``hot early universe'' of SBBC, both pure hydrogen and light nuclides can be synthesized in ECO photosphere and atmosphere. At the same time ECOs also accrete preexisting gas from the ISM. Thus a stellar mass ECO acts as the fundamental churning pot of cosmic matter: Let us describe the evolution of galactic components in a broader perspective:

It is known that star forming clouds can generate, (i) Very Massive Stars, M>10 $M_*$, (ii) Massive Stars 3$M_*$<M<10 $M_*$, (iii) Light Stars 3$M_*$<M <0.1 $M_*$ and (iv) Brown Dwarfs, M<0.1 $M_*$. And these various products would broadly evolve in the following manner:

1. Very Massive Stars Collapse to become ECOs → Eventually withers into ISM.

2. Massive Stars Collapse to become NSs and eject stellar material as well as heavy elements by Supernova explosion. Eventually NSs accrete to become ECOs → Merge into ISM.

3. Light Stars collapse to become White Dwarfs (WD) and eject material by planetary Nebulae. CO white dwarfs accrete to acquire Chandrasekhar Mass limit, then they explode completely by Type 1 SN → i.e., withers into ISM

4. In an infinitely old universe, even low mass White Dwarfs, Brown Dwarfs, Black Dwarfs have enough time to attain Chandrasekhar mass limit by accretion. Then they too eventually undergo gravitational collapse by either WD or ECO route → i.e., withers into ISM.

In an infinitely old universe, however, there will be a sort of ``steady state'' with regard to the number of various components mentioned above. And the number of light white dwarfs/brown dwarfs/black dwarfs will overwhelm the number of all other massive components because (ii) Lower mass objects/stars are preferentially formed during star formation and (ii) such low mass objects would require much larger accretion times to undergo gravitational collapse. Thus it is quite likely that a major part of the so-called ``Dark Matter'' could be in the form of light White/Brown/Dark dwarfs. All stellar activities, collapse processes and various high energy processes in the galaxies would generate photons and neutrinos of various energy ranges. Thus a portion of the ``Dark Matter'' should comprise neutrinos too. The net effect of all galactic, and stellar processes would be to irretrievably convert ``mass'' into photons and neutrinos.

It is likely that the luminous cosmic ``atom'' is immersed in a primordial tenuous cloud of cold neutral hydrogen of infinite extent and a fractal structure of dimension 1.0. If so, the luminous ``atom'' and its sub-components replenish their loss of mass (due to emission of radiation) by accreting from the infinite halo. Thus, the atom and its surrounding halo may be and churning mass into energy. This process might continue indefinitely because $\infty/\infty$ could be a finite number as well as another $\infty$ too. And, similarly, even though a given patch may be of infinite extent, it may be separated from its neighbours by infinitely larger infinities because, again, $\infty/\infty$ could be a finite number as well as another $\infty$.

### 7. Summary:

We developed here a new paradigm of an universe with nested infinities. The observed finite luminous universe is only a speckle inside a cold primordial neutral hydrogen cloud of infinite extent having a fractal dimension $D=1.0$. Though this primordial cloud has infinite mass, its mean density tapers off to zero by virtue of the fractal structure. This is possible because one might have $\infty/\infty = 0$. In the luminous patch, the accelerated expansion of the galaxies is a result of momentum conservation and loss of mass due to radiation. To honor the Copernican principle even such an infinite structure cannot be the entire ``universe'' because of inherent inhomogeneities and hierarchies. On the other hand, we postulate that such a structure is just an ``atom'' of the ``perfect'' cosmic fluid which is truly isotropic and homogeneous. Further, since the mean density of the FRW fluid must be zero, the universe comprises infinite of such ``atoms'' separated from each other by larger spatial infinities. Each ``atom'' of this fluid is a radiationless ``dust'' even though some finite parts of the ``atom'' are radiating. The grand cosmic fluid could be in a state of rest in which case Hubble expansion will have significance only within the luminous patches of the dust. And if this cosmic fluid itself would be undergoing Hubble flow, it would be a genuine case of ``space expansion''. Such a broader space expansion, at present epoch (t=$\infty$), if at all, must be with a finite speed because gravitational force is nil and mean radiation loss/unit mass too is nil. One might qualitatively understand the initial acceleration to be infinitesimally small and subsequently vanishing. It is also likely that Hubble expansion is only some dynamical effect for the luminous galaxies and the infinitely separated cosmic ``atoms'' are at rest for ever in a universe which is in a steady state except for the evolving luminous patches.

In this infinite universe, it is likely that there are equal number of matter and antimatter ``atoms'' so that the net baryon number is zero. So must be the total lepton number. And it might be also possible to infer why the total mass energy of the universe must be exactly zero.

In the absence of monotonous deceleration, the usual Raychoudhuri analysis breaks down and Eq.(9) indeed admits of an infinitely old universe even when $\dot{S} = finite$. Such an infinite universe really cannot be extrapolated back to infinite past to explore a ``shell focusing'' singularity because there would be finite mean density as soon as one would be back to S(t)=finite stage. The FRW model with inherent mean ρ =0 would then break down and one must consider effects of radiation and dissipation to discover the initial accelerating phase. Particularly, since such an infinite universe may contain equal number of matter and antimatter ``atoms'', at any finite scale size (in infinite past), it may turn into a ball of pure incoherent electromagnetic radiation for sufficiently small S(t). Then its EOS would be a strict $p = \rho c^2/3$. If so, Eq.(13) would tell that $\rho_{AGMD} = 0$ and consequently $M = 0$! Is it the reason the relativistic total mass energy of FRW model has been found to be zero by many authors?

In the present epoch with $S(t = \infty) = \infty$, the dynamics of a luminous patch however could be qualitatively similar to the standard FRW story except that any phase with a given redshift $z$ would be much older. The extreme case is that while $z = \infty$ corresponds to a finite past with $t \sim -1/H_0$ in SBBC, in EEIU, it would be $t = -\infty$. Despite being infinitely old, $\rho_{patch} > 0$ because it is possible to have $\infty/\infty = finite$ too. Eq.(8) would suggest that the properties of the observable universe would depend not only on time but on the position of the test galaxy too. It may be noted that Nan & Schwarz[24] have found that the effective EOS depends not only on time but also on scale.

Finally, the microwave background radiation here is of no primordial origin. In a related paper, we have shown that the redshifted black body temperature of an ECO, as seen by a distant observer, is unique and independent of ECO mass[25]. Further it has been shown that this temperature cannot exceed few Kelvins and thus it is justified to fix it at 2.73 K. This microwave radiation emanating from nearest massive ECO, Sgr A[*] would be scattered by hot ISM electrons and dusts and thus the point source would appear as an extended diffused one. Such a microwave haze of angular width $20^0$ has indeed discovered in the direction of Sgr A[*] [26].

And the bolometric luminosity of Sgr A[*] predicted by ECO model is
$$L_{Sgr} \approx 3\omega^{2/3} \times 10^{36} M_6^{4/3} erg/s \qquad (15)$$
where $M_6$ is the ECO mass in units of $10^6$ solar mass and the opacity related uncertainty factor $1.0 > \omega > 0.1$. Note, the observed luminosity of galactic microwave ``haze'' is [26]

$L_{WMAP} \sim (1.0 -5.0) \times 10^{36}$ erg/s ,  61 GHz>ν>23 Ghz  (16)

And total bolometric luminosity must be higher by a factor of few. Since $M_6$ could be as large as 4.0, clearly, this observed luminosity nicely matches with ECO theory prediction. All the known & unknown ECOs residing in the galaxy, particularly those close to the solar orbit would contribute their microwave mite. Accordingly as Earth would move along the ecliptic its position in the galaxy would change and there could be additional minor fluctuations in the supposed universal background. Probably such additional fluctuations/anisotropies have already been detected by WMAP [2,3, 27,28,29,30].

**Acknowledgement:** The author thanks Stanley Robertson for a careful reading of this manuscript and for encouragement. The organizers are thanked for partial financial help.